\documentstyle[aasms4]{article}
\begin{document}
\title{Dark matter concentration in the galactic center}
\author{David Tsiklauri and Raoul D. Viollier}
\affil{Physics Department, University
of Cape Town, Rondebosch 7700, South Africa}

\begin{abstract}
It is shown that the matter concentration observed through
stellar motion at the galactic center
(Eckart \& Genzel, 1997, MNRAS, 284, 576 and
Genzel et al., 1996, ApJ, 472, 153) is consistent with a supermassive
object of $2.5 \times 10^6$  solar masses composed of
self-gravitating, degenerate heavy neutrinos,
as an alternative to the black hole interpretation.
According to the observational
data,  the lower bounds on possible neutrino masses
are $m_\nu \geq 12.0$ keV$/c^2$
for $g=2$ or $m_\nu \geq 14.3$ keV$/c^2$ for $g=1$,
where $g$ is the spin degeneracy factor.
The advantage of this scenario is that it
could naturally explain the low X-ray and
gamma ray activity of Sgr A$^*$, i.e. the so called
"blackness problem" of the galactic center.
\end{abstract}

\section{Introduction}

The idea that some of the galactic nuclei
are powered by matter accretion onto supermassive black holes is based
on strong theoretical arguments (Salpeter 1964; Zel'dovich 1964;
Lynden-Bell 1969, 1978; Lynden-Bell \& Rees 1971; see Blandford
\& Rees 1992 for reviews) and observation of rapid time
variability of the emitted radiation
which implies relativistic compactness of the radiating object.
However, so far, there is no compelling proof that supermassive
black holes actually do exist, as the spatial resolution of current
observations is larger than $10^5$ Schwarzschild radii.
The standard routine in the investigation of the nature of the
dark mass distribution
at the centers of active galaxies is to observe
stellar and gas dynamics. However, gas dynamics is usually regarded
as less conclusive since it is responsive to non-gravitational
forces such as e.g. magnetic fields.

As an alternative to the black hole scenario, Moffat (1997) considered
general relativistic models
of stellar clusters with large redshifts, and he investigated whether
such objects are long-lived enough from the point of view
of evaporation and collision timescales and stability criteria.
He then showed that, in certain cases,
stellar clusters with masses $\geq 10^6$ $M_\odot$
could mimic the behavior of supermassive black holes.
A good unprejudiced review on the subject can be found in the
paper by Kormendy \& Richstone 1995.

The identification of a central supermassive object in the Milky Way
has been a source of continuous debate in the literature.
The crucial issue is whether the center of our galaxy
harbors a supermassive black hole or any other compact dark matter
object.
Theoretical papers, which are exclusively devoted to
the black hole explanation of the galactic center,
include Lynden-Bell \& Rees 1971; Rees 1987; Phinney 1989 and de
Zeeuw 1993.
However, even within this theoretical framework, there is no full
agreement: the general consensus is that the supermassive black hole
should have mass $\sim 10^6$ $M_\odot$, while
Ozernoy 1992 and Mastichiadis \& Ozernoy 1994 argue that the black
hole mass can be as low as $\sim 10^3$ $M_\odot$.
The main motivation for a $\sim 10^3$ $M_\odot$ black hole
at the galactic center  is that it would emit less
X-rays and gamma-rays than a $\sim 10^6$ $M_\odot$ black hole,
   behaving basically like a scaled-down active galactic nucleus.
   Although earlier observations (Gehrels \& Tueller 1993;
   Watson et al. 1981; Skinner et al. 1987; Hertz \& Grindlay 1984;
   Pavlinsky, Grebnev \& Sunyaev 1994)
   have shown that the central region actually does emit X-rays and
   gamma-rays,
   the supposed true center, usually assumed to be
   Sgr A$^*$, does not emit strongly,
    at least up to energies of 30 keV (Skinner et al. 1987;
    Hertz \& Grindlay 1984;
  Pavlinsky, Grebnev \& Sunyaev 1994).
   Sgr A$^*$ radiation emission data at higher energies
   have been presented by Goldwurm et al. 1994.
   They find no source associated with Sgr A$^*$, and the inferred upper
   bound  implies that the hard X-ray luminosity of Sgr A$^*$
   is a factor of $4 \times 10^7$  less than that expected for a black
   hole of $\sim 10^6$ $M_\odot$ accreting matter at the maximum stable
   rate.
   An unavoidable source of accretion is the wind from IRS 16, a nearby
   group of hot, massive stars. Since the density and velocity of the
   accreting matter are known from observations, the accretion rate is
   basically a function of the assumed black hole mass only.
   This value represents a reliable lower limit to a real rate, given the
   other possible sources of accreting matter. Based on this and on the
   theories about shock acceleration in active galactic nuclei, 
     Mastichiadis \& Ozernoy 1994  have
   estimated the expected production of relativistic particles and their
   hard radiation. Comparing their results with available X-ray and
   gamma-ray
   observations which show that Sgr A$^*$ has a relatively low activity
   level, the authors conclude tentatively that an assumed black hole in the
   galactic center cannot have a mass greater than approximately $6 \times
   10^3$ $M_\odot$. 

Other scenarios to explain low X-ray and gamma-ray activity of Sgr A$^*$
include so-called advection dominated models (Narayan, Yi \&
Mahadevan 1995;
Narayan et al., 1997; Mahadevan, Narayan \& Krolik, 1997)
which can live with  a $\sim 10^6$ $M_\odot$ black hole that
accretes matter at a realistic accretion rate of $\dot{M} \sim 10^{-5}$
$M_\odot$
yr$^{-1}$. However, most of the energy released by viscosity is carried along
with the gas and lost into the black hole, while only a small fraction
is actually radiated.

The purpose of this paper is to present an alternative model,
based on the
idea that the dark matter concentration in the galactic center
could be a ball of degenerate, self-gravitating heavy neutrinos,
which is consistent with present observational data.

\section{The model}

In the recent past, Viollier et al. have argued that
massive, self-gravitating, degenerate neutrinos arranged in balls where
the degeneracy pressure compensates self-gravity, can
form long-lived configurations that could mimic
the properties of dark matter
at the centers of galaxies
(Viollier, 1994; Viollier et al., 1993; Viollier et al., 1992).
Tsiklauri \& Viollier (1996)  demonstrated that a neutrino ball
could play a similar role as a stellar cluster in the 3C 273 quasar,
revealing its presence through the infrared bump in the emitted
spectrum.
Tsiklauri \& Viollier (1998a) further investigated the
formation  and time evolution of neutrino balls via
two competing processes:
annihilation of the particle-antiparticle pairs via weak interaction
and spherical (Bondi) accretion of these particles.
Bili\'c \& Viollier (1997a) showed how the neutrino balls could
form via a first-order phase transition
of a system of self-gravitating neutrinos in the presence
of a large radiation density background, based on the
Thomas-Fermi model at finite temperature. They find
that, by cooling a non-degenerate gas of massive
neutrinos below a certain critical temperature, a condensed phase
emerges, consisting of quasi-degenerate supermassive neutrino
balls.
General relativistic effects in the study of the gravitational phase
transition in the framework of the Thomas-Fermi model at finite temperature
were taken into account in Bili\'c \& Viollier (1997b).
A theorem was proved by Bili\'c \&
Viollier (1997c) which in brief states that the extremization of the
free energy functional of the system of self-gravitating fermions,
described by the general relativistic Thomas-Fermi model, is equivalent
to solving Einstein's field equations.

The basic equations which govern the structure of
cold  neutrino balls have been derived in the series of papers
(Viollier, 1994; Viollier et al., 1993; Viollier et al., 1992
and Tsiklauri \& Viollier, 1996); here
we adopt the notations of Tsiklauri and  Viollier (1996).
In this notation the enclosed mass of the neutrinos and antineutrinos
within a radius $r=r_n \xi$ of a neutrino ball is given by
$$
M(\xi)= 8 \pi \rho_c r_n^3  \left({-\xi^2 {{d \theta(\xi)}\over{d \xi}}
 }\right)
\equiv 8 \pi \rho_c r_n^3 \left({-\xi^2 \theta^{\prime}}\right), \eqno(1)
$$
where, $\theta (\xi)$ is the standard solution of the Lane-Emden equation
with polytropic index $3/2$, $r_n$ is the Lane-Emden unit of length
and $\rho_c$ is the central density of the neutrino ball.
In this paper we use  the length-scale 1 pc
instead of $r_n$, resulting in a trivial re-scaling of the standard
Lane-Emden equation.

To model the mass distribution usually the first moment of the collisionless
Boltzmann equation (also referred to as Jeans equation) is used
(Binney \& Tremaine, 1987)
$$
GM(R)/R= v_{\rm rot}(R)^2 - \sigma_r(R)^2
{\left({ {{d \ln n(R)}\over{d \ln R}} +
{{d \ln \sigma_r(R)^2}\over{d \ln R}}   }\right)}, \eqno(2)
$$
where $n(R)$ is the spherically symmetric space density distribution of
stars, $M(R)$ is the total included mass, $\sigma_r(R)$ is the non-projected
radial velocity dispersion and $v_{\rm rot}$ is the rotational contribution.
In order to apply Eq.(2) to the observational data, one should
relate the intrinsic velocity dispersion to the projected one via the
following Abel integrals
$$
\Sigma(p)=2 \int_p^\infty n(R)R dR/ \sqrt{R^2-p^2} \eqno(3a)
$$
$$
\Sigma(p) \sigma_r(p)^2=2 \int_p^\infty \sigma_r(R)^2
n(R)R dR/ \sqrt{R^2-p^2}, \eqno(3b)
$$
where $\Sigma(p)$ denotes surface density and $p$ is the projected
distance.
One further needs  some parametrization for $\sigma_r(R)$ and $n(R)$,
and after numerical integration of Eqs.(3), the free parameters
appearing in
$\sigma_r(R)$ and $n(R)$ should be varied in order to obtain
the best fit of
$\sigma_r(p)$ and $\Sigma(p)$ with the observational data.
Following  Genzel et al. 1996, we use parametrization
$$
n(R)={{(\Sigma_0/R_0)}\over{1+(R/R_0)^\alpha}} \eqno(4)
$$
as a model for $\Sigma(p)$. $R_0$ is related with the core radius
through $R_{\rm core}=b(\alpha)R_0$, where $b$=2.19 for $\alpha=1.8$.
Genzel et al. 1996 find that the best fit parameters for the
stellar cluster are a central density of
$\rho (R=0)=4 \times 10^6 M_\odot/{\rm pc}^3$
and a core radius of $R_{\rm core}=0.38$ pc. Thus,
for the mass distribution, Genzel et al. 1996 obtain
a black hole of $2.5 \times 10^6 M_\odot$ plus a stellar cluster
with the abovementioned physical parameters.
As mentioned earlier, we argue here that a neutrino ball composed of
self-gravitating, degenerate neutrinos within a certain mass range
could mimic the role of a black hole. This can be seen
in Fig.1, where the mass distribution of the neutrino ball
(using the rescaled Eq.(1)), with a
neutrino mass in the range of 10-25 keV$/c^2$ for $g=1$ and 2,
plus the stellar cluster is plotted. For comparison, the
$2.5 \times 10^6 M_\odot$ black hole plus  stellar cluster and pure
stellar cluster
are also shown. We gather from the graph that in the case of $m_\nu=12.013$
keV$/c^2$ for $g=2$ and $m_\nu=14.285$ keV$/c^2$ for $g=1$ (note that
these two
curves actually do overlap) the mass distribution is marginally
consistent with the
observational data. 
It is clear that for larger neutrino masses
(with corresponding degeneracy factor $g$ and with the same total mass),
the neutrino
ball would be more compact, therefore
also consistent with the observational data.
It is worthwhile to note that precise values of masses of the
neutrinos are essential since the radius of the neutrino ball, which
actually sets the neutrino mass constraints,
scales as $\propto m_\nu^{8/3}$.
To investigate what an impact the replacement of the black hole by a
neutrino ball would have,
we also calculated $\sigma_r(p)$ for both mass distributions:
a $2.5 \times 10^6 M_\odot$ black hole plus a stellar cluster and
a neutrino ball composed of $m_\nu=12.0$ keV$/c^2$
for $g=2$ or $m_\nu=14.3$ keV$/c^2$ for $g=1$ neutrinos
with the same total mass plus stellar cluster.
First we fitted the observational data taken from Eckart \& Genzel 1997,
Genzel et al. 1996 and references therein via numerical integration of the
following  expression for $\sigma_r(R)$
$$
\sigma_r(R)^2=\sigma(\infty)^2+ \sigma(2 '')^2
(R/2 '')^{-2 \beta}
$$
using Abel integrals Eq.(3). For the fit parameters we obtain
$\sigma(\infty)=59$ km/sec, $\sigma(2 '')=350$ km/sec and $\beta=0.95$,
and for the distance to Sgr A$^*$ we took 8 kpc.
The resulting $\sigma_r(p)$'s for both mass distributions are plotted
in  Fig. 2., which shows that the difference is rather
small. It is worthwhile to point out that the actual fit parameters do
not play an important role, since the aim of the graph is to demonstrate that
the substitution of the $2.5 \times 10^6 M_\odot$ black hole
by a neutrino ball  of the same mass, which is composed of self-gravitating,
degenerate neutrinos with masses of  $m_\nu = 12.0$ keV$/c^2$
for $g=2$ or $m_\nu = 14.3$ keV$/c^2$ for $g=1$,
produces a very tiny effect in the
projected velocity dispersion.  Only further
theoretical input (e.g. the use of Jeans equation) makes it possible
to discriminate between different density distribution models.
In fact, our results are in accordance with  the similar
conclusion by McGregor et al. 1996, where the authors
calculated the projected velocity dispersions by integrating the Jeans equation
with enclosed mass profiles that combine the Saha et al. 1996 mass model with
 inward extrapolation with $M(r) \propto r$ and central black holes of masses
 of $(0 - 1.5) \times 10^6 M_\odot$.

\section{Conclusions}

We have shown that a neutrino ball of total mass $2.5 \times 10^6 M_\odot$
which is composed of self-gravitating,
degenerate neutrinos and antineutrinos of mass  $m_\nu \geq 12.0$ keV$/c^2$
for $g=2$ or $m_\nu \geq 14.3$ keV$/c^2$ for $g=1$,  surrounded by a
stellar cluster
with a central density of $\rho (R=0)=4 \times 10^6 M_\odot /{\rm pc}^3$
and a core radius of $R_{\rm core}=0.38$ pc, is consistent with
 the currently available
observational data. A neutrino ball with the abovementioned
physical parameters would be virtually indistinguishable
from a black hole with the same mass,  as far as the current
observational data is concerned.

Many models were  put forward to explain the low X-ray and
gamma ray emission of the Sgr A$^*$.
Another possible solution to this "blackness problem" could be the
presence
of a neutrino ball which is  consistent with
current observational data, instead of the supermassive black hole.
In fact, in the neutrino ball scenario,
the accreting matter would experience
a much shallower gravitational potential, and
therefore less viscous torque would be exerted.
The radius of a neutrino ball of total mass
$2.5 \times 10^6 M_\odot$,
which is composed of self-gravitating,
degenerate neutrinos and antineutrinos of mass  $m_\nu = 12.0$ keV$/c^2$
for $g=2$ or $m_\nu = 14.3$ keV$/c^2$ for $g=1$,  is $1.06 \times 10^5$
larger than the Schwarzschild radius of a black hole of the same
mass. In this context, it is important to note that the accretion
radius $R_{\rm A}=2GM/v^2_{\rm w}$ for the neutrino ball,
where $v_{\rm w}\simeq 700$ km/sec is the velocity of the wind from
the IRS 16
stars, is approximately 0.02pc (Coker \& Melia, 1997), which is slightly
less than the radius of the neutrino ball, i.e. 0.02545 pc
(for $m_\nu = 12.0$ keV$/c^2$
for $g=2$ or $m_\nu = 14.3$ keV$/c^2$ for $g=1$).  The accretion radius
is the characteristic distance from the center within which the
matter is actually gravitationally captured. Therefore, in
the neutrino ball scenario, the captured accreting matter will
always experience a
gravitational pull from a mass less than the total mass of the ball.
We do not discuss  this issue  any further,
since the direct comparison of the emitted X-ray
spectra with the black hole or with neutrino ball instead
would require to go into details of current models of
X-ray emission from a compact object. The ultimate goal
of this paper was to demonstrate that our model of the mass distribution at
at the galactic center is consistent with the current observational data.

It is worthwhile to note that a possible way to distinguish
between the supermassive black hole and
neutrino ball scenarios is to track a single star, which is moving
on a bound orbit inside the radius of the neutrino ball,
over a significant part of the orbiting period.
The star trajectory, in general,
 would be an open path between the classical turning points
 $r_{\rm min}$ and $r_{\rm max}$.
The trajectory would be closed only in the case of
$1/r$ (black hole) and $r^2$ (uniform density distribution) potentials.
In the case of the black hole, the star would orbit on an ellipse,
with the black hole located at the focus, whereas
in the case of a uniform density distribution,
the center of the ellipse would
coincide with center of ball. In the
neutrino ball scenario, the trajectory of a star would be somewhat
intermediate  between the black hole and uniform density orbits.
The period of a star on an elliptical
orbit around the black hole
is  $T=2 \pi  \sqrt{a^3/G M}$, where
$a$ is the semi-major axis of the ellipse. Putting
 $2a\approx 2.545 \times 10^{-2}$ pc, with $2a$ being
the radius of the neutrino ball for $m_\nu = 12.0$ keV$/c^2$
for $g=2$ or $m_\nu = 14.3$ keV$/c^2$ for $g=1$,  and
$M=2.5 \times 10^6 M_\odot$ we thus obtain $T=85.1$ yr.
The difference between the black hole and neutrino ball
scenarios is that in the case of a neutrino ball
the period will remain roughly constant
for any orbit within the neutrino ball, as it is
well represented by an extended object with uniform
density distribution with average density about $1/6$ of the
actual central density of the neutrino ball (Viollier, 1994),
while in the black hole scenario $T$ would scale as $T \propto a^{3/2}$.
In summary, future stellar proper motion studies
on an appreciable fraction of this time-scale may by practical in
discriminating between the two scenarios.

Another possible characteristic signature of a neutrino ball at
the galactic center would be the X-ray emission line at the energy
$\sim m_\nu c^2/2$ which has a width of about the Fermi energy
($\varepsilon_F=p^2_F/2 m_\nu= (6 \pi^2/g)^{2/3}
(\hbar^2/2 m_\nu) n_\nu^{2/3}  $).
This X-ray emission, a direct consequence of the
standard electroweak interaction theory, is due to
the decay of the heavy neutrino into a photon and massless neutrino
species, both with energies $\sim m_\nu c^2/2$ (Viollier, 1994).
For Dirac neutrinos, this would generate a luminosity of
$$
L_\gamma=2.27 \times 10^{31}\left({{m_\nu c^2}\over{17.2
{\rm keV}}}\right)^5
|U_{\tau \nu_\tau} U^*_{\tau \nu_i} |^2 {{M_\nu}\over{M_\odot}}
\,\,\, {\rm erg/sec},
$$
where $U_{\tau \nu_i}$ denotes the
Cabibbo-Kobayashi-Maskawa
matrix element and $M_\nu$ is the mass of neutrino ball. Thus, putting
$M_\nu=2.5 \times 10^6 M_\odot$  and the experimental upper limit
$|U_{\tau \nu_\tau} U^*_{\tau \nu_i} |^2 \leq 10^{-3}$,
we obtain
$L_\gamma \leq 1.45 \times 10^{34}$ erg/sec.

The galactic center has been observed in the 2-10 keV range
by Koyama et al. (1996). They find that
the X-ray flux from inside the Sgr A shell (an oval region of
$\simeq 2^\prime \times 3^\prime$) is approximately $10^{-10}$ erg
cm$^{-2}$ sec$^{-1}$ in the 2-10 keV band. After correcting
for the observed absorption by a column of approximately
$7\times10^{22}$ H-atom cm$^{-2}$, they obtain a
luminosity of $\simeq 10^{36}$ erg/sec for an assumed distance of 8.5 kpc
to the galactic center.
To detect  X-rays emitted by the
neutrino ball a much higher angular resolution is needed. It would
suffice to make observations of $0.6^{\prime \prime} \times
0.6^{\prime \prime}$
(about the size of the neutrino ball) region around
Sgr A$^*$. The diffuse luminosity expected from an area corresponding
to the area of the neutrino ball
 would be $(0.6^{\prime \prime} \times
0.6^{\prime \prime})/(2^\prime \times 3^\prime)
\times 10^{36}$ erg/sec $ \approx 1.67 \times 10^{31}$ erg/sec. This number
could be even lower since it includes contributions from all energies
from 2 to 10 keV. Thus, it seems possible to detect X-ray line
of $E_\gamma \geq 6.0$ keV ($g=2$) or $E_\gamma \geq 7.1$ keV ($g=1$)
due to the radiative decay of the neutrino in the neutrino ball.
However, it might be that the
energy of the emitted X-rays is too close to the fluorescent iron
lines to be detected with the current energy resolution of CCD cameras and
gas-imaging spectrometers.

We would like to emphasize that the idea that Sgr A$^*$
may be an extended object rather than a supermassive black hole
is not new (see e.g. Haller et al., 1996; Sanders, 1992).
 To our knowledge all previous such  models  assume that the extended
object is of a baryonic nature, e.g.  a very compact stellar
cluster. However, it is commonly accepted that these models 
face problems with stability and it has been questioned whether
such clusters are 
long-lived enough, based on evaporation and collision time-scales 
stability criteria (for different point of view see Moffat, 1997).
It is interesting to note that in the context of a different object,
the center of the NGC 4258 galaxy, based on the similar criteria,
Maoz (1995) has shown (the similar study has been
performed later also for our Galaxy and other galaxies (Maoz, 1998))
that an object composed of elementary particles
would not contradict  with the observational data,
which agrees with our conclusions.

Finally, we would like to make a comment on the
neutrino mass  necessary for our model to work.
We are particularly interested in neutrinos with masses between
10 and 25 keV$/c^2$, as these could form supermassive, degenerate
neutrino balls which may explain, without invoking the black hole
hypothesis, some of the features observed around the supermassive
compact dark objects of masses ranging from $10^{6.5} M_\odot$ to
$10^{9.5} M_\odot$, which have been reported to exist at the
centers of a number of galaxies (Kormendy \& Richstone, 1995)
including our own (Genzel, Hollenbach \& Townes, 1994; Eckart \&
Genzel, 1996; Morris, 1996; Tsiklauri \& Viollier, 1998a, 1998b)
and quasi-stellar objects. A 10 to 25 keV$/c^2$  neutrino
is neither in conflict with particle and nuclear physics nor with
astrophysical observations (Viollier, 1994). On contrary, if the
conclusion of the LSND collaboration which claims to have detected
${\bar \nu}_\mu \to {\bar \nu}_e$ flavor oscillations
(Athanassopoulos et al., 1996)  is confirmed, and the quadratic
see-saw mechanism involving the up, charm and top quarks
(Gell-Mann et al., 1979; Yanagida, 1979), is the correct mechanism
for neutrino mass generation, the $\nu_\tau$ mass may very well
be within the cosmologically forbidden range between 6 and 32
keV$/c^2$ (Bili\'c \& Viollier 1997d). It is well known that
such a quasi-stable neutrino would lead to an early matter dominated
phase, which may have started as early as a few weeks after the Big-Bang.
As a direct consequence of this, the Universe would have reached the
current microwave background temperature  much too
early to accommodate the oldest stars in globular clusters,
cosmochronology and the Hubble expansion age. It is conceivable,
however,
that, in the presence of such heavy neutrinos, the early Universe
might have evolved quite differently than described in the Standard
Model of Cosmology (Kolb \& Turner, 1990, 1991;
B\"orner, 1988). Neutrino balls might have been formed in local
condensation process during a gravitational phase transition,
shortly after the neutrino matter dominated epoch began.
The latent heat produced in such a first-order phase transition,
apart from reheating the gaseous phase, might have reheated
the radiation background as well. Annihilation of the heavy
neutrinos into light neutrinos via the $Z^0$ boson
will occur efficiently in the interior
of the neutrino balls, as the annihilation
rate is proportional to the square of the number density,
which is order of $10^{25}$ particles per cm$^3$ at the center of a
few $10^9 M_\odot$ neutrino ball.
Both these processes will decrease the
contribution of the heavy neutrinos to the critical density today
and therefore, increase the age of the Universe (Kolb \& Turner, 1991).
Thus, a quasi-stable neutrino in the mass range between 10 and 25
keV$/c^2$ is not excluded by astrophysical arguments (Viollier, 1994).
In fact, it has been established in the framework
of the Thomas-Fermi model at finite temperature (Bili\'c \& Viollier 1997a)
that such neutrino balls can form via a first-order gravitational
phase transition, although, the mechanism through which the
latent heat is released during the phase transition
and dissipated into observable or perhaps unobservable matter or
radiation
remains to be identified. At this stage, however, it is still not clear
whether an energy dissipation mechanism can be found
within the minimal extension of the Standard Model of Particle
Physics or whether new physics is required in the right-handed
neutrino sector, in order to generate this efficient cooling
of the neutrino matter.

\newpage
\centerline{\bf Figure captions}
Fig.1.  Different models for the enclosed mass:
 neutrino balls
(using rescaled Eq.(1)) with the
neutrino mass in the range of 10-25 keV$/c^2$ for $g=1$ and 2
plus the stellar cluster;
$2.5 \times 10^6 M_\odot$ black hole plus the stellar cluster;
stellar cluster only. Note that mass curves for
  $m_\nu = 12.013$ keV$/c^2$
for $g=2$ and $m_\nu = 14.285$ keV$/c^2$ for $g=1$
which go through the innermost error bar do overlap.
Data points are taken from Eckart \& Genzel, 1997 and
Genzel et al., 1996 and references therein.

Fig.2. Projected velocity dispersions for different mass
models:
$2.5 \times 10^6 M_\odot$ black hole plus the stellar cluster;
neutrino ball with the same total mass and
with  $m_\nu = 12.013$ keV$/c^2$
for $g=2$ or $m_\nu = 14.285$ keV$/c^2$ for $g=1$
plus the stellar cluster. Note the tiny difference
between these two curves, as  emphasized
in the inserted window which is a zoomed region
around the innermost error bar.
The data points are taken from Eckart \& Genzel, 1997 and
Genzel et al., 1996 and references therein.
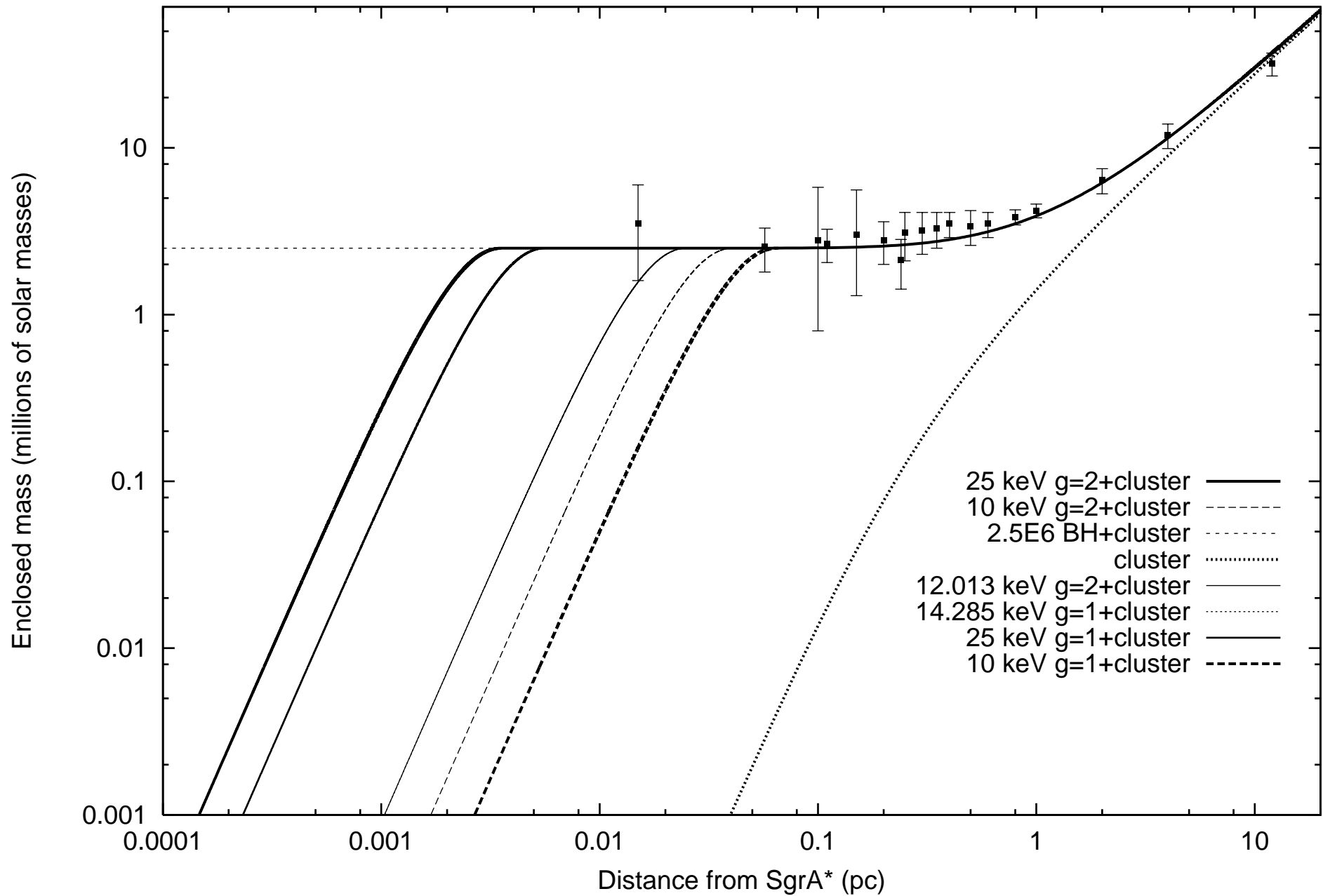
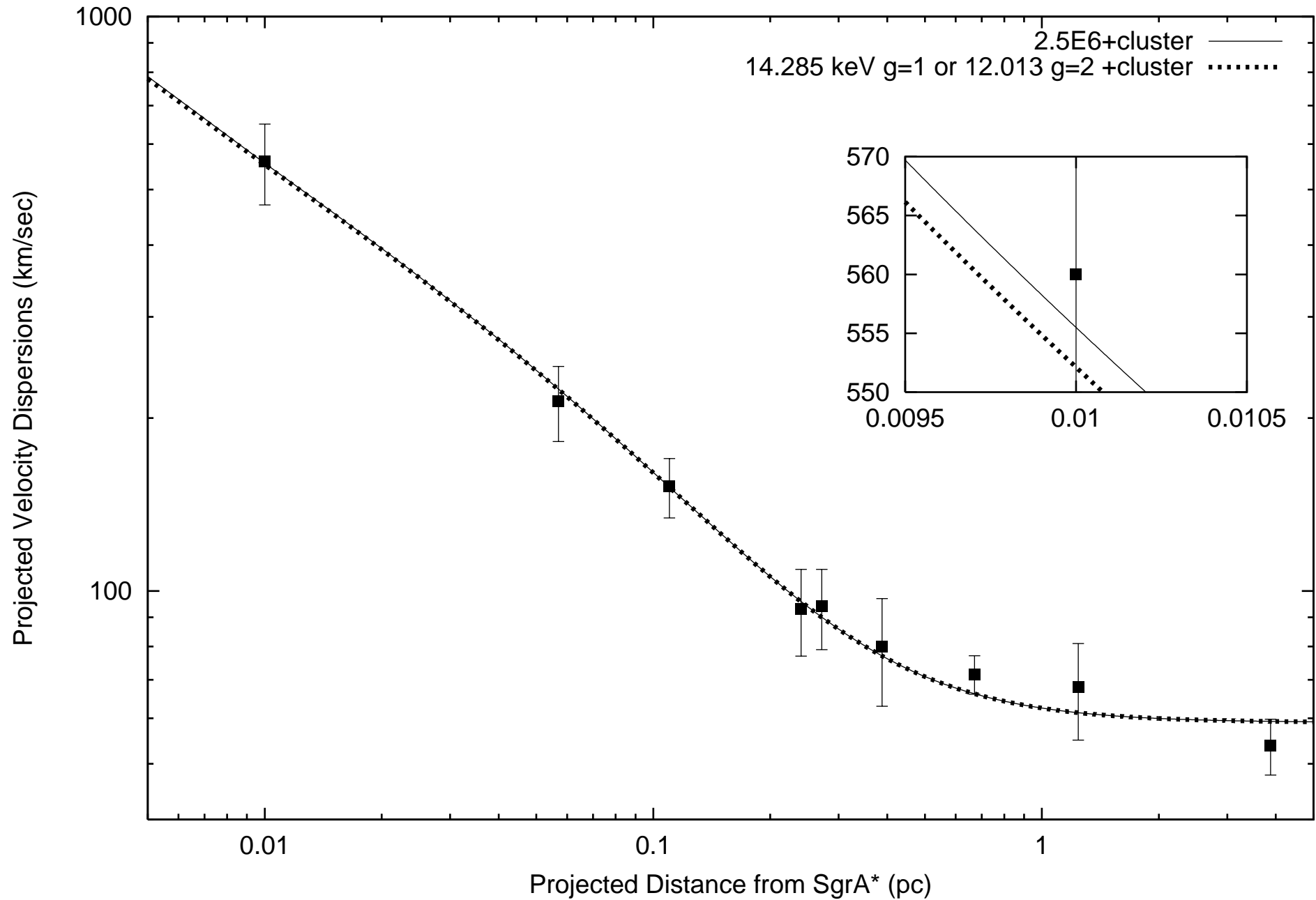
\end{document}